\documentstyle[12pt,psfig]{article}
\def\bb{\bar \beta}

\def\be{\begin{equation}}
\def\ee{\end{equation}}

\def\lsim{\raise0.3ex\hbox{$<$\kern-0.75em\raise-1.1ex\hbox{$\sim$}}}
\def\gsim{\raise0.3ex\hbox{$>$\kern-0.75em\raise-1.1ex\hbox{$\sim$}}}

\def\NP{{ Nucl.\ Phys.\ }}
\def\PL{{ Phys.\ Lett.\ }}
\def\PR{{ Phys.\ Rev.\ }}

\def\PRL{{ Phys.\ Rev.\ Lett.\ }}

\def\ZP{{ Z.\ Phys.\ }}

\begin{document}

\noindent October 31, 1999~ \hfill rev. 21.\ 1.\ 2000

\vskip 1.5 cm

\centerline{\large{\bf Polyakov Loop Percolation and}}

\medskip

\centerline{\large{\bf Deconfinement in SU(2) Gauge Theory}}

\vskip 1.0cm

\centerline{\bf Santo Fortunato and Helmut Satz}

\bigskip

\centerline{Fakult\"at f\"ur Physik, Universit\"at Bielefeld}
\par
\centerline{D-33501 Bielefeld, Germany}

\vskip 1.0cm

\noindent

\centerline{\bf Abstract:}

\medskip

The deconfinement transition in $SU(2)$ gauge theory and the
magnetization transition in the Ising model belong to the same
universality class. The critical behaviour of the Ising model 
can be characterized either as spontaneous breaking of the $Z_2$
symmetry of spin states or as percolation of appropriately 
defined spin clusters. We show that deconfinement in $SU(2)$
gauge theory can be specified as percolation of Polyakov loop clusters
with Fortuin-Kasteleyn bond weights, leading to the same (Onsager) 
critical exponents as the conventional order-disorder description 
based on the Polykov loop expectation value.

\vskip 1cm

Colour deconfinement is a well-defined phase transition in finite
temperature $SU(2)$ gauge theory; the expectation value of the
Polyakov loop serves as an order parameter determining the onset of a
spontaneous breaking of a global $Z_2$ symmetry \cite{Larry,Kuti}.
The resulting critical behaviour belongs to the universality class of
the Ising model, conjectured on the basis of effective theories
\cite{Janos,S&Y} and confirmed by lattice studies \cite{Engels}.
In full QCD, dynamical quarks act as (small) external field; hence
the conventional formalism of spontaneous symmetry breaking does not
define an order parameter, and it is not clear if deconfinement remains
a genuine phase transition. It therefore seems helpful to consider an
alternative approach to critical behaviour in the Ising model, which may
be more readily generalizable to full QCD.

\par

The magnetization transition in the Ising model (in the absence of an
external magnetic field) can be described either as the spontaneous
breaking of the $Z_2$ symmetry of the theory by spin states or as
percolation of cluster states appropriately defined in terms of the 
basic spin-spin interaction \cite{F&K}. More specifically, 
the partition function on a
two-dimensional lattice of $L^2$ sites is conventionally defined as sum
over all possible spin states $\{s\}\equiv\{s_i
=\pm1~\forall~i=1,...,L^2\}$,
\be
Z(T) = \sum_{\{s\}} b(\{s\}) \label{1}
\ee
with the Boltzmann weight
\be
b = \exp\{ (J/T) \sum_{\langle i,j \rangle} s_is_j\},\label{2}
\ee
where $T$ is the temperature and $J$ the spin-spin
coupling strength; the sum $\langle i,j \rangle$ runs over next
neighbour spin pairs, with $i<j$.  
Equivalently, it can be written \cite{Baxter} as sum over all
$2^E$ clusters states $G$, specified in terms of $l$ bonds forming $C$
connected clusters on a lattice consisting of $E$ links,
\be
Z(T) = \sum_{G} 2^C v^l. \label{3}
\ee
where 
\be
v = (\exp\{2(J/T)\}-1). \label{4}
\ee
defines the bond weight. Note that Eqs.\ (\ref{1}) and (\ref{3}) differ
by a T-dependent factor related to the zero in energy.

\par

In the spin formulation, the order parameter is given by the spontaneous
magnetization $m(T)$, defined as the average spin per lattice site,
\be
m(T) = \langle |s| \rangle \equiv { \sum_{\{s\}} \{ |\sum_i s_i|/L^2 \}~
b(\{s\}) \over \sum_{\{s\}} b(\{s\}) }.\label{5}
\ee
Its behaviour near the critical point is governed by the critical
exponent $\beta$, with
\be
m(T) \sim (T_c-T)^{\beta}, ~~~~T\lsim T_c; \label{6}
\ee
the divergence of the corresponding susceptibility is with
\be
\chi_m(T) \sim \langle s^2 \rangle - \langle |s| \rangle^2 \sim
|T-T_c|^{-\gamma}. \label{7}
\ee
determined by the exponent $\gamma$.

\par

In the percolation formulation, the order parameter becomes the
percolation strength $P(T)$, defined as the probability that a randomly
chosen site in the thermodynamic limit belongs to an infinite cluster,
\be
P(T) \sim (T_c - T)^{\beta}, ~~~~T\lsim T_c.\label{8}
\ee
The corresponding susceptibility is the average cluster size
$S(T)$, excluding percolating clusters, and its divergence is
governed by
\be
S(T) \sim |T-T_c|^{-\gamma} \label{9}
\ee
in the vicinity of the critical point.

Spin and cluster formulations thus provide two equivalent ways to
specify the critical behaviour of the Ising model. While the spin
version is based on the onset of spontaneous breaking of the global
$Z_2$ symmetry of the Ising Hamiltonian, the cluster version uses the
onset of percolation of clusters whose Fortuin-Kasteleyn bond weights
are also determined by Ising dynamics.

\par

Since for a fixed space dimension $d$, the critical behaviour of the
Ising model and of finite temperature $SU(2)$ gauge theory are in the
same universality class, one may expect that deconfinement can also
be formulated as percolation. The aim of this paper is to show that for
$d=2$ and for a specific lattice regularization, this is indeed the
case. The generalization to $d=3$ appears straight-forward and is in
progress.

\par

To implement the cluster formulation, we use the droplet approach
introduced by Coniglio and Klein \cite{C&K}. For the Ising model, this
method generates equilibrium configurations using the Boltzmann weights
and then defines clusters as regions of parallel spins connected by
bonds, using Fortuin-Kasteleyn bond weights
\be
p=1-\exp\{-2(J/T)\}. \label{9a}
\ee
The use of both Boltzmann
and bond weights can be avoided, and there exist implementations using
the Fortuin-Kasteleyn bond weights only \cite{HMO} - \cite{Wolff}. In our
context, the Coniglio-Klein implementation seems preferable mainly for
computational reasons.

\par

In the Ising model, clusters are thus defined as regions of parallel
spins connected by bonds, with the probability for bonding given by 
Eq.\ (\ref{9a}). In two-dimensional $SU(2)$ lattice gauge theory, the
underlying manifold becomes a $N_s^2 \times N_t$ lattice, with $N_s$
sites in each spatial and $N_t$ sites in the temperature direction. The
Ising spins $s_i=\pm 1$ at sites $i=1,...,L^2$ are replaced by Polyakov
loops $L_i$,
\be
L_i \sim \left\{ {\rm Tr} \prod_{\tau=1}^{N_t}
U_{(i;\tau,\tau+1)} \right\}, \label{10}
\ee
where $U_{(i;\tau,\tau+1)}$ are $SU(2)$ matrices associated to the 
link at spatial site $i$ connecting the temporal planes $\tau$ and
$\tau+1$.

The matrix product in Eq.\ ({\ref{10}) becomes a loop closed by
periodicity in the temperature direction. We thus replace the discrete
spin values $s_i=\pm 1$ by spins of continuous size $L_i=\pm |L_i|$ at
each lattice site $i$. Clusters are now defined as regions of like-sign
Polyakov loops (say $L_i \geq 0$) connected by bonds distributed
according to the bond weight
\be
p_{i,j} = 1 - \exp\{-2\kappa L_i L_j\}, \label{11}
\ee
where $\kappa$ is determined by the dynamics of $SU(2)$ gauge theory;
it corresponds to the $J/T$ in the Ising weight Eq.\ (\ref{9c}).

\par

The identification of the underlying dynamics as $SU(2)$ gauge theory
is contained in the $\kappa$ in Eq.\ (\ref{11}). A general solution for
this does not seem to exist so far. In the context of universality
studies \cite{Janos,S&Y}}, the $SU(2)$ action is formally reduced to an
effective action in terms of the Polyakov loop variables $L_i$, so it
should be possible to identify $\kappa$ in such a framework. We shall
here make use of strong coupling calculations \cite{G&K}, in which
\be
\kappa \simeq (\bb/4)^2 \label{12}
\ee
was found to provide a good approximation (90\% accuracy) for $N_t=2$; 
here $\bb=(4/g^2)$ denotes the coupling parameter in the $SU(2)$ 
lattice action.

\par

With this, the Polyakov loop percolation problem is fully defined, and
we proceed with a lattice study of the percolation strength $P(\bb)$ and
the cluster size $S(\bb)$. Our analysis is based on four sets of data on
$N_s^2 \times 2$ lattices, with $N_s$=64, 96, 128 and 160. The
simulations were carried out on workstations for $N_s \leq 128$ and
on a Cray T3E (ZAM, J\"ulich) for $N_s$=160. The updates consist of one
heatbath and two overrelaxation steps. For the $64^2\times 2$ and
$96^2\times 2$ lattices we evaluated configurations every six updates,
for $128^2\times 2$ and $160^2\times 2$ every eight updates, measuring
in each case $P(\bb)$ and $S(\bb)$. The expectation value of the
Polyakov loop gives $\bb_c(L) \simeq 3.464$ as the critical
point on $N_s^2\times 2$ lattices in the limit $N_s \to \infty$
\cite{Teper}, giving us an idea of the range to be studied.

\par

\begin{figure}[htb]
\centerline{\psfig{file=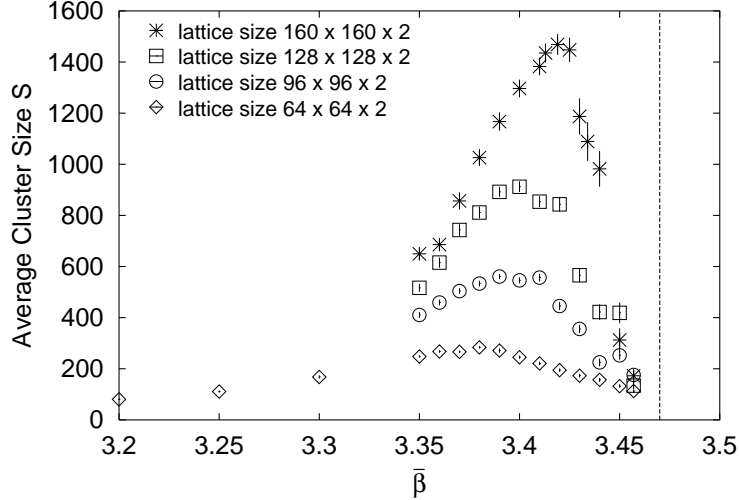,height= 70mm}\hspace{1.2cm}}
\vspace{-0.2cm}
\caption{Cluster size as function of $\bb$ near the critical point
$\bb_c(L)$ (dashed line).}
\end{figure}

\begin{figure}[htb]

\centerline{\psfig{file=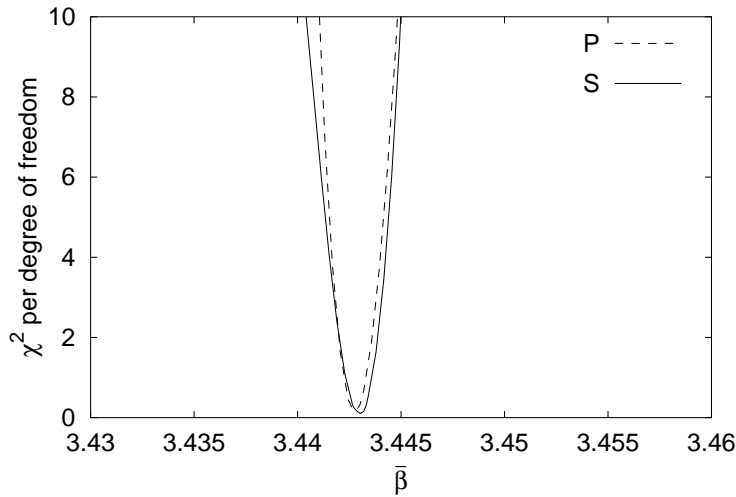,height= 70mm}}
\vspace{-0.2cm}
\caption{$\chi^2$ distributions for scaling fits of 
$P(\bb)$ and $S(\bb)$.}
\end{figure}

A first scan for values $3.1 \leq \bb \leq 3.5$ leads to the behaviour
of $S(\bb)$ shown in Fig.\ 1. It is seen that $S(\bb)$ peaks slightly
below $\bb_c(L)$; with increasing $N_s$, the peak moves towards $\bb_c(L)$
and the peak height increases.

\par

In a second step, we carried out high-statistics simulations in a
narrower range $3.410 \leq \bb \leq 3.457$ around the transition. In
general, we performed between 30000 and 55000 measurements per $\bb$
value, with the higher number taken in the region of the interval
closest to the eventual critical point. To obtain the behaviour in 
the limit $N_s \to \infty$, we applied the density of states method (DSM)
\cite{DSM}. This method generates for each lattice size
by interpolation further values of $P(\bb)$ and $S(\bb)$. 
The percolation critical point $\bb_c(P)$ and the critical exponents 
are then determined through the $N_s$-dependence of the observables. At
$\bb_c(P)$, fits of the form $P(\bb) \sim N_s^{-\beta/\nu}$ or
$S(\bb) \sim N_s^{\gamma/\nu}$ should lead to a minimal $\chi^2$ \cite{EMSZ}.
Fig.\ 2 shows the $\chi^2$/d.f.\ resulting from fits of $\log P$ and 
$\log S$ vs.\ $\log~N_s$ at each value of $\bb$ in the range $3.43 \leq 
\bb \leq 3.46$. The two $\chi^2$ curves show pronounced minima with 
remarkable overlap.

\par

In Table 1 we show the results for 95\% confidence level, comparing the
critical $\bb_c(P)$ for Polyakov loop percolation to the $\bb_c(L)$
from spontaneous symmetry breaking \cite{Teper},
and the percolation exponents to the
Onsager values for the two-dimensional Ising model ($\beta$=1/8,
$\gamma$=7/4, $\nu$=1). The critical exponents are seen to be in
excellent agreement. Since $\nu$ is determined by taking derivatives
of the interpolated curves of $P(\bb)$ and $S(\bb)$ with respect to $\bb$,
the precision of its determination is reduced in comparison to that
of $\beta/\nu$ and $\gamma/\nu$. 
The critical values of $\bb$ are very close,
but they do not overlap within errors. In view of the approximate
solution for $\kappa(\bb)$ used here, small deviations are not
unexpected. Monte Carlo renormalisation group techniques may well
allow a more precise determination of the bond weight (\ref{11}). \cite{MC}.

\par

\begin{center}
\begin{tabular}{|c||c|c|c|c|}
\hline
& & & & \\
& $\bb_c$ & $\beta/\nu$ & $\gamma/\nu$ & $\nu$\\
& & & & \\
\hline
\hline
& & & & \\
Percolation & 3.443$^{+0.001}_{-0.001}$ & 0.128$^{+0.003}_{-0.005}$
& 1.752$^{+0.006}_{-0.008}$ & 0.98$^{+0.07}_{-0.04}$ \\
& & & & \\
\hline
& & & & \\
Spont.\ Symm.\ Breaking & 3.464$^{+0.012}_{-0.016}$ & 0.125 & 1.75 & 1.00 \\
& & & & \\
\hline
\end{tabular}\end{center}

\centerline{Table 1: Critical parameters for $SU(2)$
lattice gauge theory with $N_t=2$.}

\bigskip

We thus conclude that for the specific case considered here,
two-dimensional $SU(2)$ lattice gauge theory with $N_t=2$, deconfinement
can indeed be specified through Polyakov loop percolation. To make this
more general, further work is necessary. The most crucial open problem
is certainly the general form of $\kappa$ in Eq.\ (\ref{11}). In
particular, it is not clear why the general $SU(2)$ action should lead
to a nearest-neighbour form, and how the `temperature' $\kappa$ can be
defined more generally, away from the strong coupling limit $N_t=2$. For
the latter problem, the use of Monte Carlo renormalisation group
techniques \cite{MCRG} may be of help. If and when these problems are
solved, the corresponding percolation study has to be
carried out for three space dimensions and for $SU(3)$ gauge theory as
well, where the transition becomes first order.
\par

Finally we return to the motivation mentioned at the beginning. For spin
systems, percolation in coordinate space remains as critical phenomenon
even in the presence of an external field \cite{Kertesz}, although this
criticality is now not connected any more to singular behaviour of the
partition function. In principle, this could provide the basis for a
deconfinement order parameter in full QCD with dynamical quarks
\cite{HS-perco}.

\bigskip

\centerline{\bf Acknowledgements}

\medskip

H.-W.\ Huang participated in the early stages of this work; we thank
him very much for his helpful contributions. Furthermore, we would like
to thank Ph.\ Blanchard, P.\ Bia{\l}as, J.\ Engels, D.\ Gandolfo and 
F.\ Karsch for stimulating discussions. The financial support 
of the EU-Network ERBFMRX-CT97-0122 and the DFG Forschergruppe Ka 1198/4-1
is gratefully acknowledged.


\begin{thebibliography}{99}

\par

\bibitem{Larry} L.\ D. McLerran and B.\ Svetitsky, \PL 98 B
(1981) 195.

\bibitem{Kuti} J.\ Kuti, J.\ Pol\'onyi and K.\ Szlach\'anyi,
\PL 98B (1981) 199.

\bibitem{Janos} J.\ Pol\'onyi and K. Szlach\'anyi, \PL 110B (1982) 395.

\bibitem{S&Y} B.\ Svetitsky and L.\ G.\ Yaffe, \NP B 210 [FS6] (1982)
423.

\bibitem{Engels} J.\ Engels et al., \PL B 365 (1996) 219.

\bibitem{F&K} C.\ M.\ Fortuin and P.\ W.\ Kasteleyn, Physica 57 (1972)
536.

\bibitem{Baxter} R.\ J.\ Baxter, S.\ B. Kelland and F.\ Y.\ Wu, J. Phys.
A 9 (1976) 397.

\bibitem{C&K} A.\ Coniglio and W. Klein, J.\ Phys.\ A 13 (1980)
2775.

\bibitem{S-W} R.\ H.\ Swendsen and J.-S.\ Wang, \PRL 58 (1987) 86.

\bibitem{HMO} H.\ Meyer-Ortmanns, \ZP 27 (1985) 553.

\bibitem{Wolff} U.\ Wolff, \PRL 62 (1989) 361.

\bibitem{G&K}  F.\ Green and F.\ Karsch, \NP B 238 (1984) 297.

\bibitem{Teper}  M.\ Teper, \PL B 313 (1993) 417.

\bibitem{DSM} M.\ Falconi et al., \PL 108 B (1982) 331; \par
E.\ Marinari, \NP B 235 (1984) 123; \par
G.\ Bhanot et al., \PL 183 B (1986) 331; \par
A.\ M.\ Ferrenberg and R.\ H.\ Swendsen, \PRL 61 (1988) 2635 and 63
(1989) 1195.

\bibitem{EMSZ} J.\ Engels et al., \PL B 365 (1996) 219.

\bibitem{MC} A.\ Goksch and M.\ Ogilvie, \PRL 54 (1985) 1772;\par
R.\ V.\ Gavai, A.\ Goksch and M.\ Ogilvie, \PRL 56 (1986) 815;\par
M.\ Okawa, \PRL 60 (1988) 1805.

\bibitem{MCRG} A.\ Gonzales-Arroyo and M.\ Okawa, \PR D 35 (1987) 672:\par
M.\ Okawa, \PRL 60 (1988) 1805.

\bibitem{Kertesz} J.\ Kert\'esz, Physica A 161 (1989) 58.

\bibitem{HS-perco} H.\ Satz, \NP A642 (1998) 130c.

\end{thebibliography}
\end{document}